# ALTERNATIVE SAMPLE MASS MEASUREMENT TECHNIQUE FOR OSIRIS-REX SAMPLE COLLECTION PHASE

**Huikang Ma[1]**, **Michael Skeen[2]**, **Ryan Olds[1]**, **Brennen Miller[2]**, **Dante S. Lauretta[3]**

The Origins, Spectral Interpretation, Resource Identification, and Security-Regolith Explorer (OSIRIS-REx) spacecraft is the third NASA New Frontiers Program mission and arrived at the near-Earth asteroid (101955) Bennu in December 2018. Following completion of sample collection in October 2020, otherwise known as Touch-And-Go (TAG), the OSIRIS-REx spacecraft was set to verify its collected sample mass requirement (> 60g of material). The thoroughly tested Sample Mass Measurement (SMM) method was to be used for this verification. Imaging of the Touch-And-Go Sample Acquisition Mechanism (TAGSAM) was received shortly following the TAG event, intended to ensure mechanism health prior to moving forward with the SMM activity. These images displayed sample leakage, prompting discussion for alternative paths forward. Risk of continued sample loss and a desire to retain as much material as possible lead the team to pursue an accelerated sample stow schedule and forgo the planned SMM activity. Once the sample was safely stowed in the return capsule an alternative SMM method was proposed. The alternative SMM technique utilized reaction wheel momentum data from identical TAGSAM movements prior to and following the TAG event to estimate changes in spacecraft moment of inertia. Conservation of momentum was used to isolate the sample mass from this inertia change. Using this new method, the spacecraft team was able to successfully estimate collected sample mass to be $250.37 \pm 101$ g.

[1] OSIRIS-REx Guidance Navigation and Control Team, Lockheed Martin Space, Littleton, CO, 80125
[2] OSIRIS-REx Systems Team, Lockheed Martin Space, Littleton, CO, 80125
[3] OSIRIS-REx Principal Investigator, University of Arizona, Tucson, AZ, 85721

## INTRODUCTION

The Origins, Spectral Interpretations, Resource Identification, and Security-Regolith Explorer (OSIRIS-REx) mission launched from Cape Canaveral in September of 2016. With a primary mission requirement of collecting and returning a 60 gram regolith sample from asteroid Bennu (Figure 1) to Earth.

Bennu is both the most accessible carbonaceous asteroid and one of the most potentially Earth-hazardous asteroids currently known (Lauretta et al, 2015). As a B-type carbonaceous asteroid, Bennu may represent an important source of volatiles and organic matter to early Earth as well as being a direct remnant of the original building blocks of the terrestrial planets (Lauretta el al, 2017). Knowledge of the nature of near-Earth asteroids such as Bennu is fundamental to understanding planet formation and the origin of life. The return to Earth of pristine regolith samples with known geologic context enables precise analyses that cannot be duplicated by spacecraft-based instruments alone, thereby revolutionizing our understanding of the early Solar System. Study of Bennu addresses multiple NASA objectives to understand the origin of the Solar System and the origin of life and will provide a greater understanding of both the hazards and resources in near-Earth space, serving as a precursor to future missions to asteroids.

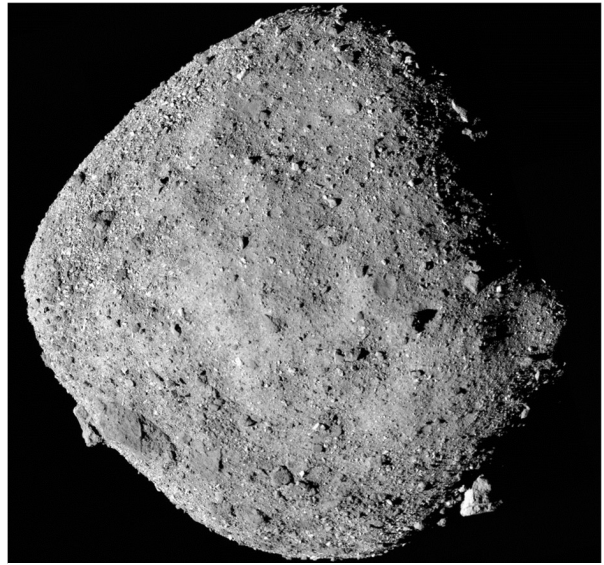

**Figure 1: Bennu PolyCam Mosaic. Credit: NASA/Goddard/University of Arizona**

## MISSION OVERVIEW

Over the course of ~2 years the OSIRIS-REx spacecraft journeyed from its Florida launch site to asteroid Bennu, a mission phase known as outbound cruise. Its arrival at Bennu in December 2018 marked the start of a new mission phase, proximity operations. OSIRIS-REx and the flight operations team spent the next year and a half imaging and studying the surface of Bennu in detail, with the goal of determining a suitable site for sample collection. The spacecraft was to descend to the asteroid surface, contacting the specified site to collect a sample (regolith). The interface between the asteroid surface and OSIRIS-REx was the Touch-and-Go Sample Acquisition Mechanism (TAGSAM), a device used to collect and capture regolith. This whole process would take a matter of seconds before the spacecraft would autonomously thrust itself away from the surface to safety. Early images of Bennu from the proximity operations phase made it clear that selecting a suitable collection site would not be an easy task. The surface of Bennu was much rougher and rockier than expected. Ground based analysis of asteroid Bennu had originally predicted a surface composed primarily of small particles, almost sand like. Throughout the following months many potential sample sites, though not ideal or within original requirements, were identified. These went through a lengthy selection process and despite the challenge a final sample site was selected, now known as "Nightingale" (Figure 2).

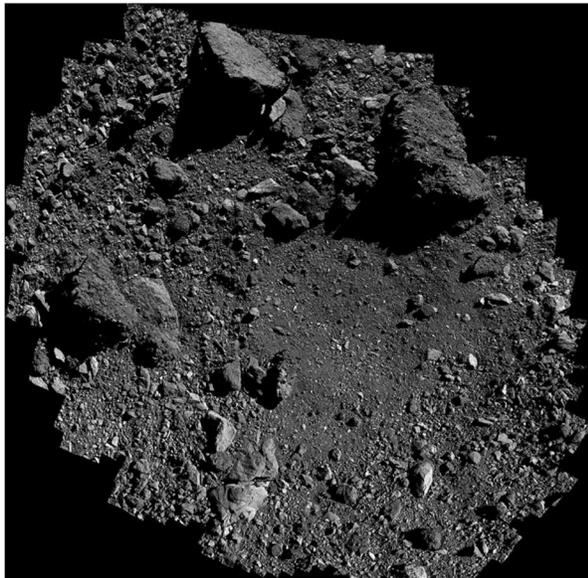

**Figure 2: Nightingale sample site. Credit: NASA/Goddard/University of Arizona**

Quickly following selection of the Nightingale site OSIRIS-REx began its Touch-and-Go (TAG), sample collection phase. OSIRIS-REx can autonomously descend to the surface of Bennu and guide itself to the Nightingale site using Natural Feature Tracking (NFT) technology. This technique was verified through multiple "dress rehearsals" in which ORISIRS-REx performed the TAG sequence of events but executed a planned waive-off of the activity at various stages in its decent to the surface. Two were completed, Checkpoint rehearsal and Matchpoint rehearsal. Each providing invaluable experience prior to executing the real event.

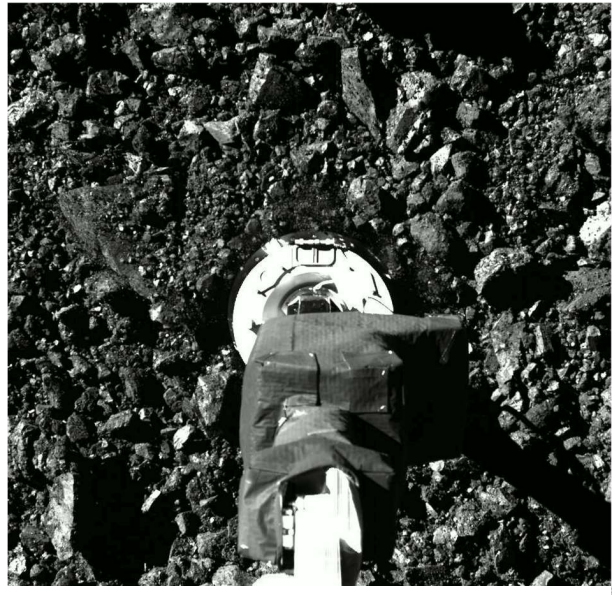

On October 20th, 2020 OSIRIS-REx successfully touched down at the Nightingale site, then using a burst of compressed nitrogen gas to force surface regolith into the TAGSAM mechanism (Figure 3). The TAG event was performed flawlessly, and initial imaging following its execution showed promising signs that a sufficient sample mass was collected. However, imaging alone did not provide a direct measurement that the 60 g pristine regolith requirement was met

**Figure 3: TAGSAM during OSIRIS-REx Sample Collection Event. Credit: NASA/Goddard/University of Arizona**

**SAMPLE MASS MEASUREMENT BACKGROUND**

An accurate estimation of collected sample mass was needed in order to verify the primary mission objective and ensure a second TAG attempt was not required. OSIRIS-REx was to use a method known as sample mass measurement (SMM) for this in-flight sample mass validation. The spacecraft would perform two sets of two inertia spins (A and B configurations shown in Figure 4), one set before TAG (#1) and another set after TAG (#2). The A configuration places the TAGSAM arm in the TAG configuration (drive both TAGSAM shoulder and elbow joint to 216.2 deg), which centers the TAGSAM head over the body Z axis of the spacecraft. The spacecraft is then spun about its Z axis, collecting data from Reaction Wheels (RWA) and Inertial Measurement Units (IMU). These data can be post processed to obtain an accurate estimate of the spacecraft's moment of inertia. In the "A" configurations inertia contribution of the TAGSAM head and its contents are minimal since there is no moment arm between the masses and the spin axis. The "B" configurations fully extend the TAGSAM arm to place the TAGSAM head as far from the Z axis as possible (drive TAGSAM shoulder joint to 90 deg and elbow joint to 180 deg), thus maximizing the inertia of the sample mass collected for the measurement. By calculating the difference between inertia estimates in the A configuration followed shortly thereafter by an estimate in the B configuration, the change in inertia created by moving the TAGSAM arm between the two configurations is isolated. Those results are then repeated following the TAG event, when the TAGSAM head contains regolith sample. Differencing those inertia differences will yield a change in inertia, which is then used to calculate collected sample mass. Many in flight rehearsals of this method were performed prior to sample collection, all resulting in sample mass results within expected error bars. The latest in flight rehearsal before TAG yielded a sample mass uncertainty of 10.7 g, 74.2% below the requirement and 55.3% below the previous current best estimate (CBE).

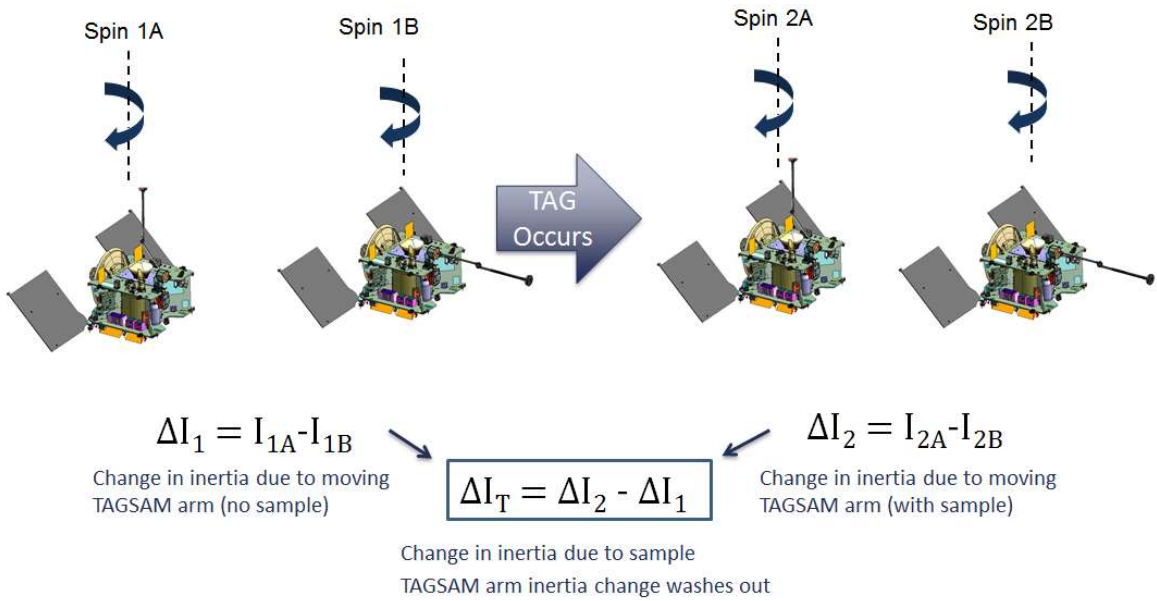

**Figure 4: Sample Mass Estimation Configurations**

## TAG RESULTS AND SAMPLE STOW

A sample imaging campaign was performed following the TAG event. Its purpose was to provide pictures of the TAGSAM head prior to proceeding with SMM, verifying no significant damage had occurred upon impact with the surface. The images received depicted something unexpected. It appeared that a large amount of sample had been collected — a reason to celebrate — but the mylar flap located below the sample chamber was being held open by some large particles thus caused sample leaking into space (Figure 5). With this knowledge, a decision was made to forgo the SMM activity, which might cause loss further loss of sample material, and proceed directly with an accelerated sample stow schedule. This would prevent any further sample from being lost to space. Performing SMM would have delayed this process and enacted additional forces on the sample head, potentially freeing up more sample to escape. The team proceeded with the accelerated sample stow activities, ultimately leading to successfully stowing the TAGSAM head and the regolith sample it contained (Figure 6).

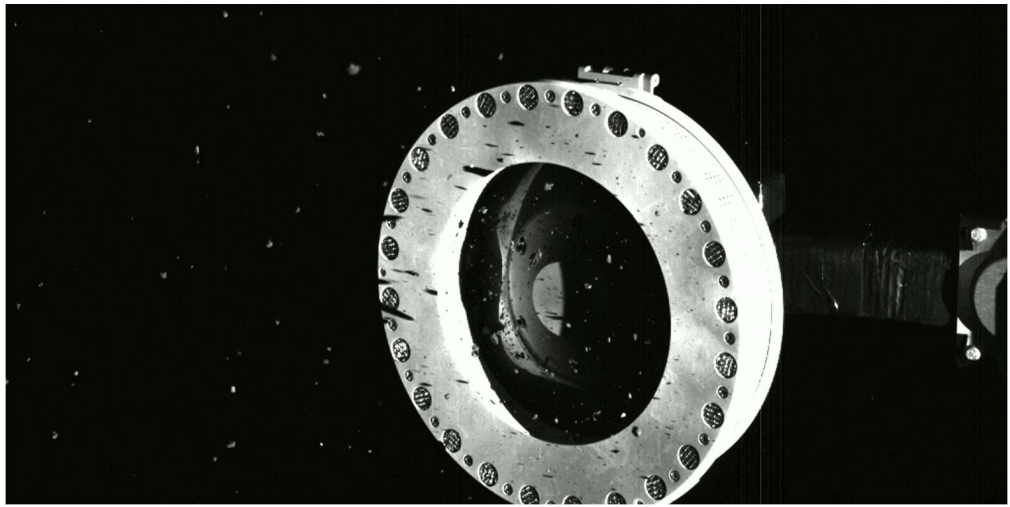

**Figure 5: Sample leaking from TAGSAM head during sample imaging campaign. Credit: NASA/Goddard/University of Arizona**

The sample was safe but without having obtained a direct measurement of collected sample mass. At this point the only estimates of collected sample mass were from visual inspection alone. Knowing this mass was important as it would determine if the primary mission requirement of 60 grams had been met. Also, Since the Sample Return Capsule's (SRC) atmospheric re-entry trajectory will vary depending on the capsule mass, estimating the sample mass is highly beneficial. Therefore, while the original measurement method was no longer an option, it would be beneficial to devise another approach to obtain a quantitative estimate. Fortunately, the guidance and navigation team at Lockheed Martin developed an alternative approach to estimate the sample mass with data already collected from previous spacecraft activities.

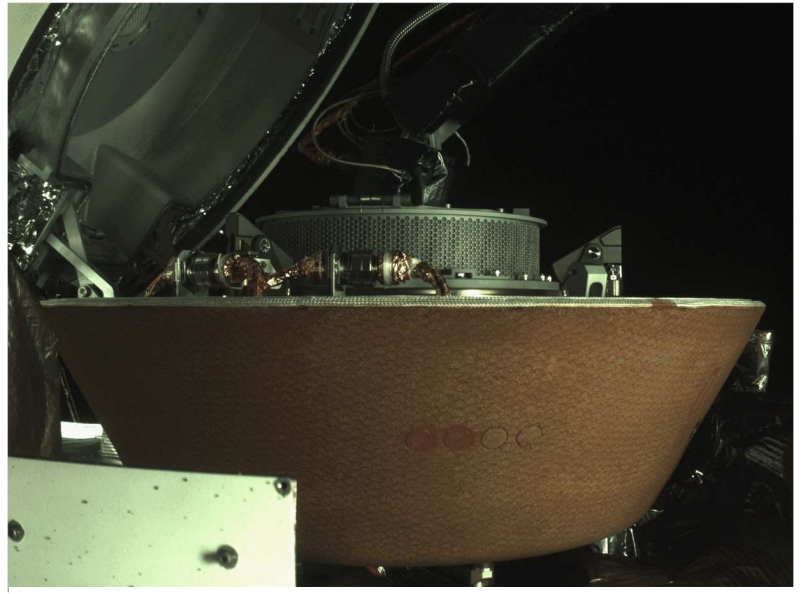

**Figure 6: Sample collection head latched into the SRC. Credit: NASA/Goddard/University of Arizona**

## ALTERNATIVE SAMPLE MASS MEASUREMENT TECHNIQUE

Instead of spinning the spacecraft along its z-axis and determining the change in the spacecraft inertia, the alternate Sample Mass Measurement technique calculates the change of the TAGSAM arm momentum while the TAGSAM arm is moving (shown in Figure 7). Two events are necessary in order to determine the sample mass: The first event was executed prior to collecting the sample (with no sample mass in the sample head) and the second event was executed after collecting the sample (with sample mass in the sample head).

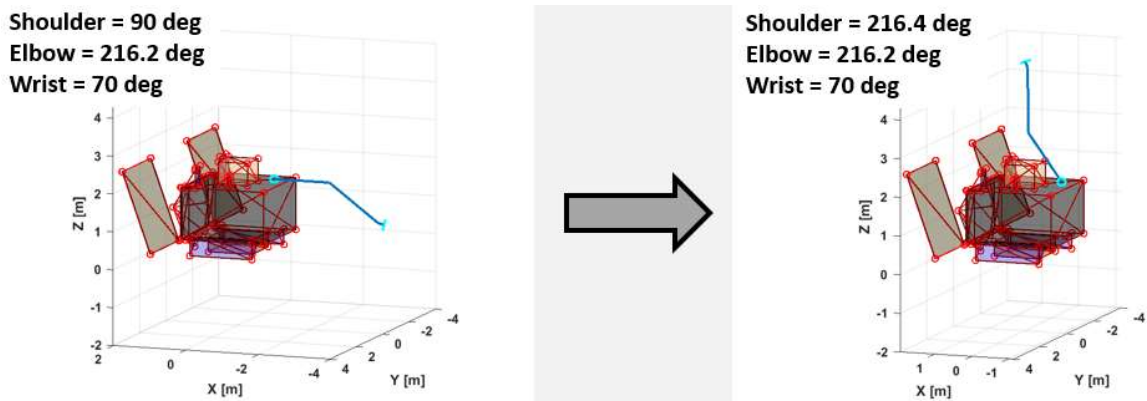

**Figure 7: Alternative Sample Mass Measurement data collection period**

The method exploits the fact that the momentum transfer from the reaction wheels to the TAGSAM arm during this motion will be different with and without the sample mass in the sample head. The conservation of momentum for each event can be expressed as follows.

$$\vec{H}_T = \vec{H}_{B0} + \vec{H}_{R0} + \vec{H}_{A0} = \vec{H}_{Bt} + \vec{H}_{Rt} + \vec{H}_{At} \qquad (1)$$

Where $\vec{H}_T$ is the total system angular momentum of the spacecraft. $\vec{H}_B$ represents the angular momentum stored in the spacecraft bus, $\vec{H}_R$ represents the angular momentum stored in the spacecraft reaction wheels, and $\vec{H}_A$ represents the angular momentum stored in the spacecraft arm. The "*0*" suffix on these variables represents the initial angular momentum prior to arm motion and the "*t*" suffix represents the angular momentum at any given time during the arm motion. Because the arm motion is performed in an inertially held attitude, body rates are negligible, so Equation 1 can be simplified by eliminating the $\vec{H}_{B0}$ and $\vec{H}_{Bt}$ terms. TAGSAM is also initially at rest so the $\vec{H}_{A0}$ is also zero. These assumptions lead to the following simplification of Equation 1.

$$\vec{H}_{Rt} - \vec{H}_{R0} = -\vec{H}_{At} \qquad (2)$$

Denoting Equation 2 with a "1" suffix for Event 1 (pre-TAG) and with a "2" suffix for Event 2 (post-TAG) and then differencing these two equations leads to Equation 3.

$$\left(\vec{H2}_{Rt} - \vec{H2}_{R0}\right) - \left(\vec{H1}_{Rt} - \vec{H1}_{R0}\right) = -\left(\vec{H2}_{At} - \vec{H1}_{At}\right) \qquad (3)$$

Equation 3 shows that the change in reaction wheel momentum observed during each individual event can be differenced to find the change in momentum stored by the robotic arm between each arm motion event. This "double-difference" in the reaction wheel momentum will therefore be equal to the difference in stored momentum caused by the sample in the TAGSAM sample head.

These Equations so far have assumed that no external torques are acting on the spacecraft during motion of the TAGSAM arm. This is not true since solar radiation pressure is acting on the spacecraft during this time. This in turn generates an external torque which the reaction wheels must counter to maintain the inertially fixed attitude of the spacecraft. The additional momentum accumulated during the activity due to solar radiation torques must be accounted for to avoid biasing the measurement. To do this, the solar radiation torque was estimated based on GN&C simulation software as well as past flight data. This torque was multiplied by time during the arm motion and the resulting momentum was subtracted from the reaction wheel momentum.

$$\vec{H}'_{Rt} = \vec{H}_{Rt} - \tau_{srp} t \qquad (4)$$

Where $\vec{H}'_{Rt}$ is the corrected reaction wheel momentum to account for solar radiation pressure torques. Note that the time *t* is the time from the start of the activity. Equation 3 then becomes:

$$\left(\vec{H2}'_{Rt} - \vec{H2}_{R0}\right) - \left(\vec{H1}'_{Rt} - \vec{H1}_{R0}\right) = -\left(\vec{H2}_{At} - \vec{H1}_{At}\right) \qquad (5)$$

The left-hand side of Equation 5 is known from telemetry as it only requires the reaction wheel speeds and the rotor inertias. The right-hand side of Equation 5, can be rewritten to expose the contribution of the sample mass which is ultimately what we want to solve for. Note that since the right-hand side is the difference in angular momentum stored by the TAGSAM arm between event 1 and event 2, it can be expressed using components of mass that changed between event 1 and event 2. The only two mass elements on the arm that changed during TAG were the collection of the sample mass and the expulsion of the nitrogen gas. Therefore, event 1 (pre-TAG) had a full gas bottle, but an empty sample head, whereas event 2 (post-TAG) had an empty gas bottle but a full sample head.

$$\vec{H2}_{At} - \vec{H1}_{At} = \vec{H2}_m - \vec{H1}_{gas} = \left(\vec{r}_m^{cg} \times m_s \vec{v}_s\right) - \left(\vec{r}_{gas}^{cg} \times m_{gas} \vec{v}_{gas}\right) \qquad (6)$$

Where $\vec{r}_m^{cg}$ is the position vector from the spacecraft c.g. to the sample mass, $m_s$ is the sample mass, $\vec{v}_s$ is the linear velocity of the sample mass as the arm moves, $\vec{r}_{gas}^{cg}$ is the position vector from the spacecraft c.g. to the gas canister, $m_{gas}$ is the mass of the expelled gas, and $\vec{v}_{gas}$ is the linear velocity of the gas as the arm moves. These quantities are depicted in the figure below.

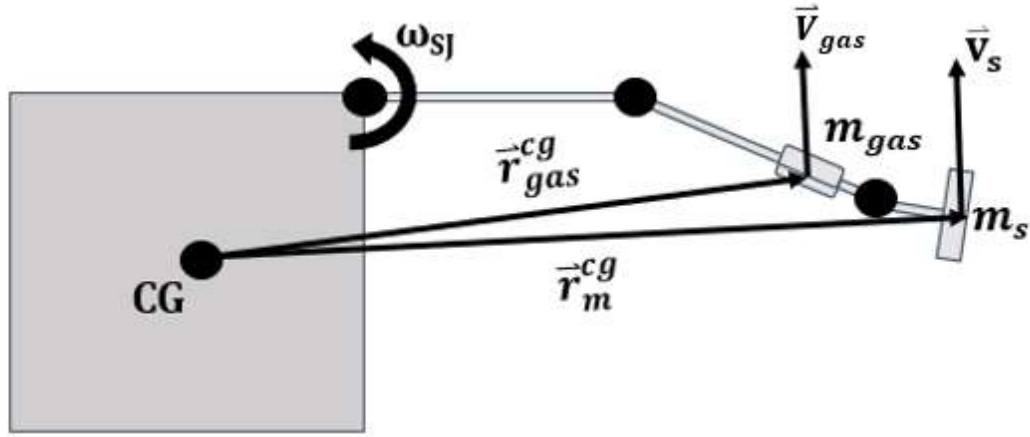

**Figure 8: TAGSAM parameters for Equation 6**

Solving for the sample mass and plugging in the known quantities on the left-hand side of Equation 5 yields:

$$m_s = \frac{\left[-(\overline{H2}'_{Rt}-\overline{H2}_{R0})+(\overline{H1}'_{Rt}-\overline{H1}_{R0})+\left(\vec{r}^{cg}_{gas}\times m_{gas}\vec{v}_{gas}\right)\right]}{\vec{r}^{cg}_m\times\vec{v}_s} \quad (7)$$

**PERTURBATIONS AND ERROR BUDGETS**

The team developed a detailed error budget to quantify the Sample Mass Measurement uncertainty, this error budget allowed the team to ensure at least 60g of sample mass had been collected to meet the mission requirement. Significant analysis was done pre-launch and refined with in-flight calibrations to quantify error sources for the baseline SMM (Skeen et al. 2019). However, these analysis tools did not directly translate to the alternative SMM method in many cases. Instead, limited analysis was done for some error sources in the budget below (Table 1), marked (A). In addition, some of the error sources were best characterized by analysis of pre-TAG in-flight motions to provide a rough assessment of CBE performance (but are not bounding), marked (F). These error sources were combined by RSS, which presumes that errors follow Gaussian distributions and are not correlated.

Table 1: Alternative Sample Mass Measurement error budget

| Subsystem | Error Sources | CBE Measurement Error (g) |
|---|---|---|
| TAGSAM | Pogo (A)<br>Sample distribution (A) | +/-10g |
|  | U-Joint and Gimbal Angle Uncertainties (A, F)<br>Shoulder motor alignment (A, F) | +/-13g |
| GN&C and Propulsion | Measurement accuracy (F)<br>Solar radiation pressure (A, F)<br>Post-slew pendulum motion (F) | +/- 100g |
| Structures & Mechanisms | Thermal distortion (A) | Negligible |
| Payloads / Ops / Other | Dust Loading (A)<br>Outgassing (A) | Negligible |
| **Total (RSS)** |  | **+/- 101g** |

**Error sources due to TAGSAM:**

The TAGSAM presented several possible sources of error. Since the sample is not constrained within the TAGSAM head, the assumed moment arm between the spacecraft CG and sample may have error. To bound this, the OSIRIS-REx Guidance Navigation and Control team performed an analysis with the sample mass location dispersed by 10 cm in worst case directions from the nominal assumed location and the results showed it contributed +/-10g of error to SMM. Likewise, the predicted location of the spacecraft CG was also dispersed by 1 cm and found to result in a negligible error to SMM.

The TAGSAM pogo compressed during contact with the asteroid, as designed. A potential source of error between pre-TAG and post-TAG measurements would arise if the pogo did not fully extend after TAG; however based on detailed image analysis of the pogo after TAG including geometric reconstruction of the TAGSAM joint positions from TAGCAMS image data, there was no observable shift in the pogo due to TAG. Likewise, there was no observable shift in the alignment of the TAGSAM shoulder motor, which would have caused a change in the angular momentum direction between pre-TAG and post-TAG articulations of the TAGSAM arm.

The TAGSAM u-joint and gimbal angle positions were determined by geometric reconstruction of joint position in TAGCAMS image data to be approximately 4 deg for the wrist between pre-TAG and post-TAG images, and approximately 3 deg for the U-joint. These joint position differences contribute error in the measured inertia between pre-TAG and post-TAG TAGSAM motions, totaling +/-13g equivalent error to the measured mass.

**Error sources due to GN&C and Propulsion:**

There are many sources of error from GN&C and propulsion that affect the alternative SMM method, including reaction wheel alignment, reaction wheel inertia uncertainty, tachometer timing, noise and filtering, attitude control errors, solar radiation pressure, TAGSAM arm tube vibration and propellant pendulum motion. As the significant effort required to build a new simulation to perform detailed analysis was not feasible in the short timelines immediately following TAG, the team elected to use previous in-flight data to provide an estimate to the magnitude of these error sources. Table 2 shown 8 pairs of in-flight data the team used to calculate the system errors. The biggest error came from case 7 (-106.27 grams of error after SRP correction), this was due to a spacecraft slew which occurred too close in time to the TAGSAM arm motion, causing excessive propellant motion resulting in larger errors.

Table 2: pre-TAG in-flight TAGSAM arm motion data used to calculate system errors

| Analysis Case: | Event 1 (S/C Attitude) | Event 2 (S/C Attitude) | Results with SRP correction [g] | Comments |
|---|---|---|---|---|
| Case 1 | SMM5 (Sun-point) | SMM6 (Sun-point) | -32.75 | SMM6 TAGSAM arm movement data had bigger than expected noise caused bigger error. |
| Case 2 | 2018SampleImg (Rearing-unicorn) | SMM5 (Sun-point) | -14.85 | |
| Case 3 | 2018SampleImg (Rearing-unicorn) | SMM6 (Sun-point) | -45.91 | SMM6 TAGSAM arm movement data had bigger than expected noise caused bigger error. |
| Case 4 | Pre-CPR SMM (Sun-point) | Pos-CPR SMM (Sun-point) | -2.65 | |
| Case 5 | CPR (Sun-point) | Pos-CPR SMM (Sun-point) | -37.81 | CPR Sun-Earth slew completed 5 min before the arm motion, fuel were not settled when the |

| | | | | TAGSAM started arm movement. |
|---|---|---|---|---|
| Case 6 | Pre-MPR SMM (Sun-point) | Pos-MPR SMM (Sun-point) | 6.19 | |
| Case 7 | MPR (Sun-point) | Pos-MPR SMM (Sun-point) | -106.27 | MPR Sun-Earth slew completed 4 min before arm motion, fuel were not settled when the TAGSAM started arm movement. |
| Case 8 | Pos-MPR SMM (Sun-point) | Pre-TAG SMM (Sun-point) | 49.72 | |

Spacecraft activities explanation:
SMM5 & SMM6: The 5th and 6th sample mass measurement calibration activity performed on 2018.
2018SampleImg: TAGSAM sample imaging rehearsal performed on 2018, in between SMM5 and SMM6.
CPR: OSIRIS-REx Check Point Rehearsal performed on April 2020.
Pre-CPR SMM & Pos-CPR SMM: Pre-CPR and Post-CPR sample mass measurement calibration activity.
MPR: OSIRIS-REx Match Point Rehearsal performed on August 2020.
Pre-MRP SMM and Pos-MPR SMM: Pre-MPR and Post-MPR sample mass measurement calibration activity.

Spacecraft attitude explanation:
Sun-Point: Sun vector is on spacecraft +X axis.
Rearing-Unicorn: Sun vector is on spacecraft +X/-Z axis.

Since the previous flight data were not sufficient to definitively separate contributions from these error sources, significant margin was added to the error observed in the worst case (after throwing out case 5 and case 7, where the propellant motion was too excessive). The resulting error allocation was chosen to be +/-101g.

As discussed in the Technique section, the solar radiation pressure acting on the vehicle created an error in calculating angular momentum change which had to be corrected. This correction was applied by using GN&C analysis tools to predict the amount and direction of angular momentum contributed by solar radiation pressure for the various sun ranges and spacecraft attitudes applicable to each motion of the TAGSAM arm used for measurement and calibration of the alternative SMM.

While a precise amount of mass error could not be generated due to propellant motion in the fuel tank, the analysis was able to predict a worst-case damping coefficient for the propellant management device (PMD) damping of fuel motion after vehicle slews. The damping of 10% with a 0.007 Hz period provided analytical evidence to remove a couple of pre-TAG calibration cases from the error analysis as the vehicle had slewed shortly before the TAGSAM arm was moved in those cases and the resulting measured errors were not applicable to the quiescent conditions used for the alternative SMM data.

**Error sources due to Structures & Mechanisms:**

A thermal analysis was performed to examine the temperature changes on the TAGSAM arm as it moves from a shadowed position behind the spacecraft into sunlight. The maximum predicted temperature delta over this motion was <10 °C, which causes a very small thermal distortion (<1 mm) and thus contributes negligible error to angular momentum and mass measurements.

**Error sources due to Payloads / Ops / Other:**

During the TAG event, significant amounts of dust and regolith material was ejected and may have collected on the TAGSAM arm. This material would appear as an increased angular momentum during TAGSAM motion; however that material would not be stored within the Sample Return Capsule and be returned to Earth. Post-TAG imagery of the TAGSAM arm did not show measurable dust loading on the visible portions of the TAGSAM arm. While some individual particles were visible in other images of the spacecraft deck, a small number of particles like these if present on the TAGSAM arm would be relatively small compared to the measured sample mass and their location (and thus moment arm for angular momentum) on the TAGSAM arm would also be less than the sample mass. Thus, the potential error due to dust and particle loading on the TAGSAM arm is considered negligible.

Outgassing had been observed in previous in-flight calibrations of the baseline SMM methodology due to the variety of surfaces exposed to the sun while performing the SMM spins, and mitigations were implemented to minimize the resulting error (Skeen et al. 2019). The alternative SMM method did not require spinning of the vehicle, and so the spacecraft surfaces exposed to the sun had ample time to outgas any volatiles that may have accumulated during the TAG event. This potential error source is also considered negligible.

**ALTERNATIVE SAMPLE MASS MEASUREMENT RESULTS**

As mentioned in the previous section, at least two TAGSAM arm movement datasets are needed (one pre-TAG and one post-TAG) in order to determine the collected sample mass, there were two TAGSAM arm movement activities performed after TAG. Therefore, the team used a total of 3 datasets to estimate the collected sample mass using the alternative sample mass measurement method. These 3 datasets were:

1. pre-TAG calibration TAGSAM arm dataset, this activity represented the spacecraft and TAGSAM behavior prior to sample collection.

2. post-TAG sample imaging campaign TAGSAM arm dataset, this activity was performed 2 days after the TAG event, it best represented the amount of sample OSIRIS-REx collected after TAG.

3. post-TAG sample stow TAGSAM arm dataset, this activity was performed one week after TAG during the sample head stow, it best represented the amount of asteroid sample OSIRIS-REx stowed into the Sample Return Capsule (SRC).

Figure 9 shown the spacecraft Y-axis angular momentum comparison during the TAGSAM arm movement between pre-TAG calibration (Event-1) and sample imaging campaign (Event-2). If there were no sample collected in the sample head, the momentum difference between this two datasets would be positive since Event-2 would have less momentum due to mass loss caused by nitrogen gas firing. However, the momentum difference during TAGSAM arm movement clearly shown it had a noticeable negative bias (Figure 9 bottom plot), indicated bigger momentum from Event-2, which means there were asteroid samples in the sample container.

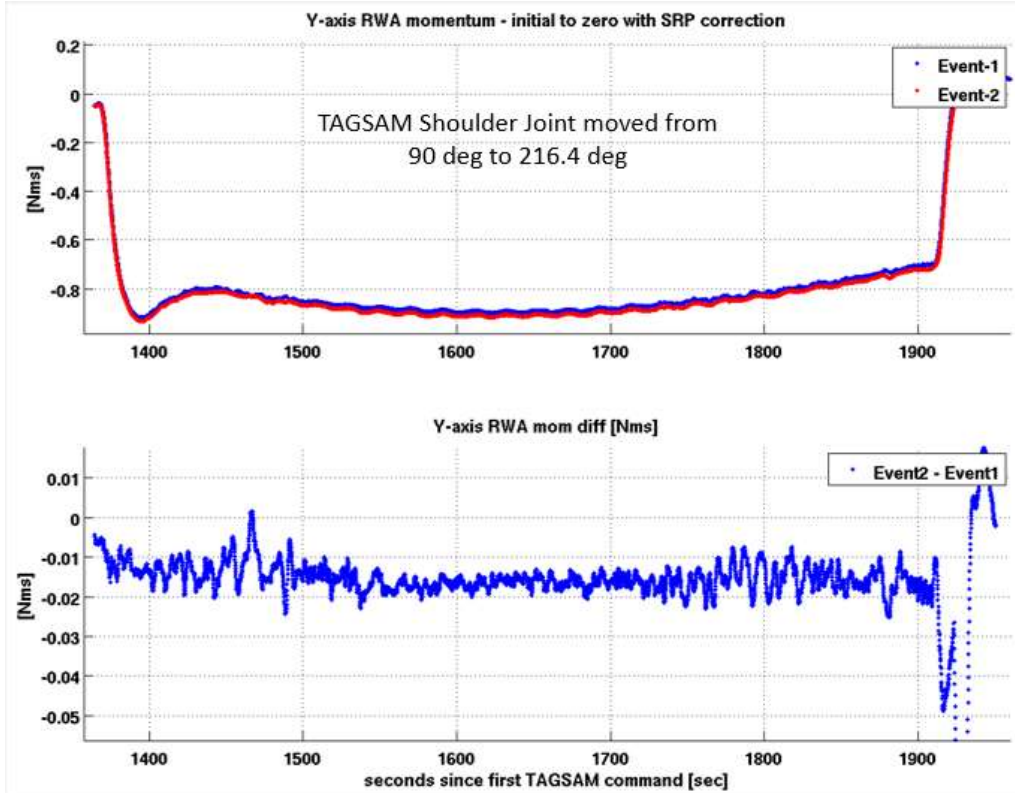

**Figure 9: Spacecraft Y-axis momentum difference during TAGSAM arm movement between pre-TAG calibration and post-TAG sample imaging campaign**

Knowing that there were momentum differences between pre-TAG and post-TAG activities, a sample mass can be estimated using equation (7) listed in the previous section. Table 3 below shown the estimated sample mass using sample imaging campaign dataset and sample stow dataset. As previous section pointed out, solar radiation pressure must be corrected in order to estimate the sample mass accurately, especially considering that pre-TAG calibration and post-TAG activities had a different attitude relative to the Sun.

Table 3: Sample mass estimation using alternative SMM method

| Event 1 (S/C Attitude) | Event 2 (S/C Attitude) | Est. Mass Difference [g] | Possible measurement error sources |
|---|---|---|---|
| pre-TAG calibration (Sun-Point) | Post-TAG Sample Imaging Campaign (Rearing-unicorn) | 326.91 | pre-TAG calibration: Reaction wheel desaturation performed ~5 min before the start of the TAGSAM arm movement, fuel might not be completely settled when arm started moving.<br><br>Post-TAG Sample Imaging Campaign: Spacecraft slew completed ~8 min before the start of the TAGSAM arm movement, fuel might not be completely settled when arm started moving. |
| pre-TAG calibration (Sun-Point) | Post-TAG Sample Stow (Rearing-unicorn) | 250.37 | pre-TAG calibration: Reaction wheel desaturation performed ~5 min before the start of the TAGSAM arm movement, fuel might not be completely settled when arm started moving. |

| | | | Post-TAG Sample Stow: Spacecraft slew completed ~8 min before the start of the TAGSAM arm movement, fuel might not be completely settled when arm started moving. |
|---|---|---|---|
| <u>Spacecraft attitude explanation:</u><br>Sun-Point: Sun vector is on the spacecraft +X axis.<br>Rearing-Unicorn: Sun vector is on the spacecraft +X/-Z axis. | | | |

The error bound for the alternative SMM method was set to be +/- 101 grams. Therefore, at the time of sample imaging campaign, the amount of sample in the TAGSAM sample head was estimated to be 326.91 +/- 101 grams, and the sample stow measurement resulting in an sample estimate of 250.37 +/- 101 grams in the sample head. Besides the measurement errors, the mass difference between this two estimations was likely caused by the sample leaked into space during the sample imaging campaign activity (Figure 5).

A separate analysis was performed using the dataset from the two post-TAG activities to estimate how much sample was leaked into space. Figure 10 shown the spacecraft Y-axis angular momentum comparison during the TAGSAM arm movement between sample imaging campaign (Event-1) and sample stow activity (Event-2). The momentum difference (Figure 10 bottom plot) shown a positive bias meaning Event-2 had less momentum compares to Event-1 which indicated a mass loss. There's no nitrogen gas firing between this two events therefor a modified version of equation 7 can be used to calculate the mass difference. It was determined that a mass difference of -67.25 grams between this two events, as pervious discussed, most of the mass loss was likely due to the sample leak during the sample imaging campaign.

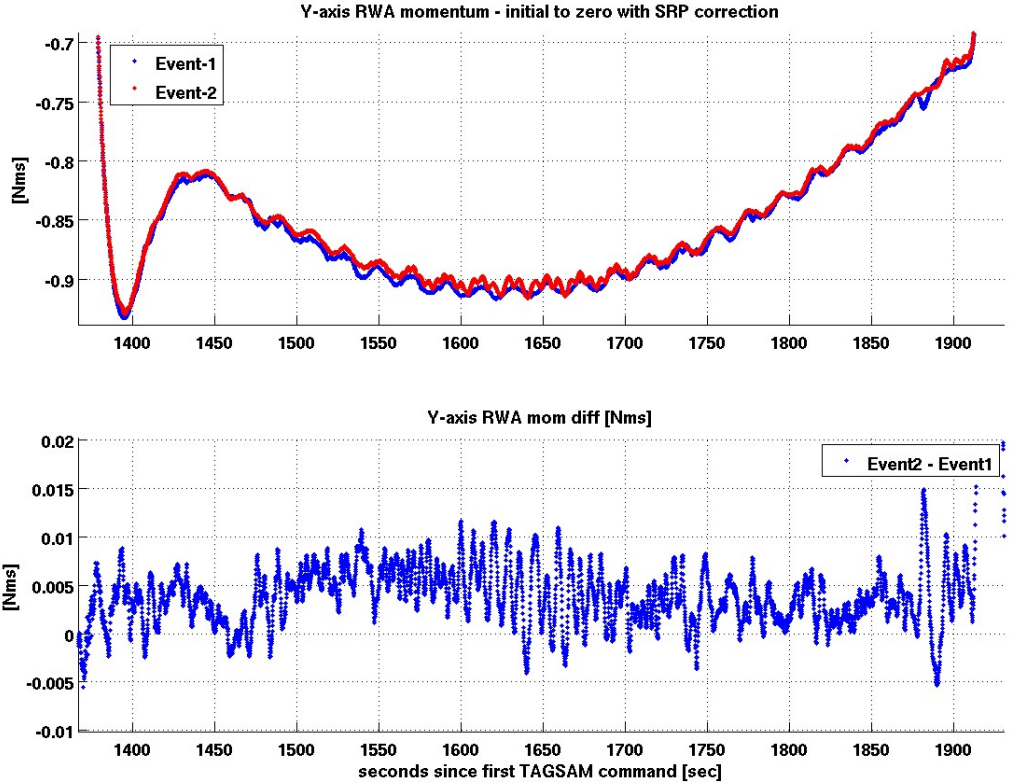

**Figure 10: Spacecraft Y-axis momentum difference during TAGSAM arm movement between post-TAG sample imaging campaign and post-TAG sample stow activity**

## CONCLUSION

The original Sample Mass Measurement activity was cancelled because the team noticed the asteroid sample was leaking into space after the TAGSAM sample imaging campaign. The spacecraft team then developed an alternative sample mass measurement method using existing data collected during the TAGSAM arm movement period to estimate the sample mass inside the sample container. The spacecraft team determined the sample container likely contained 326.91 +/- 101 gram of sample at the time of sample imaging campaign and 250.37 +/- 101 gram of sample when the sample is stowed into Sample Return Capsule (SRC). A separate analysis was performed in order to estimate how much sample was leaked during the sample imaging campaign, it was estimated 67.25 grams of sample loss after sample imaging campaign.